# A Novel P-bit-based Probabilistic Computing Approach for Solving the 3-D Protein Folding Problem

Chao Fang[1†], Yihan He[1†], Xiao Gong[1], and Gengchiau Liang[1,2*]

[1]Department of Electrical and Computer Engineering, National University of Singapore, 117583 Singapore

[2]Industry Academia Innovation School, National Yang-Ming Chiao Tung University, Hsinchu, Taiwan

†These authors contribute equally to this work

*Corresponding Author: gcliang@nycu.edu.tw

In the post-Moore era, the need for efficient solutions to non-deterministic polynomial-time (NP) problems is becoming more pressing. In this context, the Ising model implemented by the probabilistic computing systems with probabilistic bits (p-bits) has attracted attention due to the widespread availability of p-bits and support for large-scale simulations. This study marks the first work to apply probabilistic computing to tackle protein folding, a significant NP-complete problem challenge in biology. We represent proteins as sequences of hydrophobic (H) and polar (P) beads within a three-dimensional (3-D) grid and introduce a novel many-body interaction-based encoding method to map the problem onto an Ising model. Our simulations show that this approach significantly simplifies the energy landscape for short peptide sequences of six amino acids, halving the number of energy levels. Furthermore, the proposed mapping method achieves approximately 100 times acceleration for sequences consisting of ten amino acids in identifying the correct folding configuration. We predicted the optimal folding configuration for a peptide sequence of 36 amino acids by identifying the ground state. These findings highlight the unique potential of the proposed encoding method for solving protein folding and, importantly, provide new tools for solving similar NP-complete problems in biology by a probabilistic computing approach.

## I. INTRODUCTION

The protein folding problem has attracted considerable attention due to its inherent complexity and critical role in biological processes. Proteins are composed of amino acid chains that fold into specific three-dimensional (3-D) structures to exert their biological functions [1]. Protein misfolding is associated with a variety of diseases, including cancer [2], Alzheimer's disease [3], and Parkinson's disease [4]. Therefore, accurately predicting protein folding is crucial for diagnosing these diseases and understanding their underlying mechanisms [5], [6], [7]. Additionally, predicting potential protein folding configurations is also invaluable in the field of drug discovery and design [8].

Numerous simplified mathematical models have been proposed to address the protein folding problem, among which the Hydrophobic-Polar (HP) model is one of the most widely used models [9]. This model assumes all amino acids are positioned within a two-dimensional (2-D) or 3-D lattice grid and categorizes them as either hydrophobic (H) or polar (P). It emphasizes the interactions between hydrophobic amino acids to assess the stability of the protein fold and suggests that the sequence of amino acids directly dictates the folding configuration of the amino acid chain. The hypothesis has been substantiated by various experimental studies [10], [11], [12]. Despite its apparent simplicity, the protein folding problem, as simplified by the HP model, can produce qualitatively meaningful results and significantly contribute to advancing solutions. However, the protein folding problem simplified by the HP model remains inherently complex and challenging. It is classified as NP-complete, indicating that no polynomial-time algorithm can address it efficiently [13], [14]. In recent years, numerous computational approaches have been explored to address this challenge. Traditional methods such as genetic algorithms [15], ant colony optimization [16], and hill-climbing algorithms [17] have shown promise in solving the HP model for small-scale protein folding problems. However, these techniques encounter significant challenges related to computational time and energy efficiency when applied to long-chain protein folding.

In non-conventional computational fields, on the other hand, researchers have attempted to utilize quantum computing to address the protein folding problem by mapping it onto a Hamiltonian, ensuring that the optimal folding configuration is mapped to the ground state[18], [19]. An early effort involved mapping the 2-D HP protein folding problem onto a Hamiltonian encoding both path and grid positions [20]. Subsequently, new encoding methods were developed to improve performance, which map the same problem with the "Turn" encoding method [21], encoding the turns of the path rather than the path itself. The method was further extended to 3D cases and applied to other models [22], [23]. However, the "Turn" encoding method presents several issues. First, the algorithms used to encode the problem into a Hamiltonian based on the "turn" method often exhibit exponential time complexity, and some of them are not polynomial-time algorithms. It takes significantly more time as the sequence length increases. Furthermore, As the sequence extends, with an increasing number of turns, numerous many-body interactions emerge. This requires additional auxiliary nodes to create a quadratic Hamiltonian and divides the large problem into multiple sub-problems, complicating the processing of longer amino acid sequences. Later, an improved encoding way is proposed, which encodes both path and grid positions while splitting odd and even nodes to simplify the Hamiltonian [24], [25]. However, this mapping approach remains limited to 2-D scenarios. In addition, due to the inherent limitation of the number of qubits in quantum computing, the application scale of these quantum-computing-based models is constrained to small-scale problems.

In this study, we present the protein folding solver based on the probabilistic circuit (p-circuit) with p-bits for the first time. Compared to quantum computing, probabilistic computing systems have flexible hardware implementation and can operate at room temperature [26], which makes it feasible to construct large-scale Ising machines in practical p-circuits for solving protein folding problems. We also propose a method for mapping the 3-D HP model-based protein folding problem onto the Ising model. Our mapping method simplifies the mapping process compared to conventional turn-based methods, significantly enhancing efficiency when solving large-scale protein folding problems. Additionally, we improve the performance of the mapped Ising model by incorporating many-body interactions. We provide a more concise energy landscape by utilizing many-body interactions while carefully controlling the number of many-body connections to maintain p-circuit complexity, thereby greatly reducing the difficulty in the annealing process.

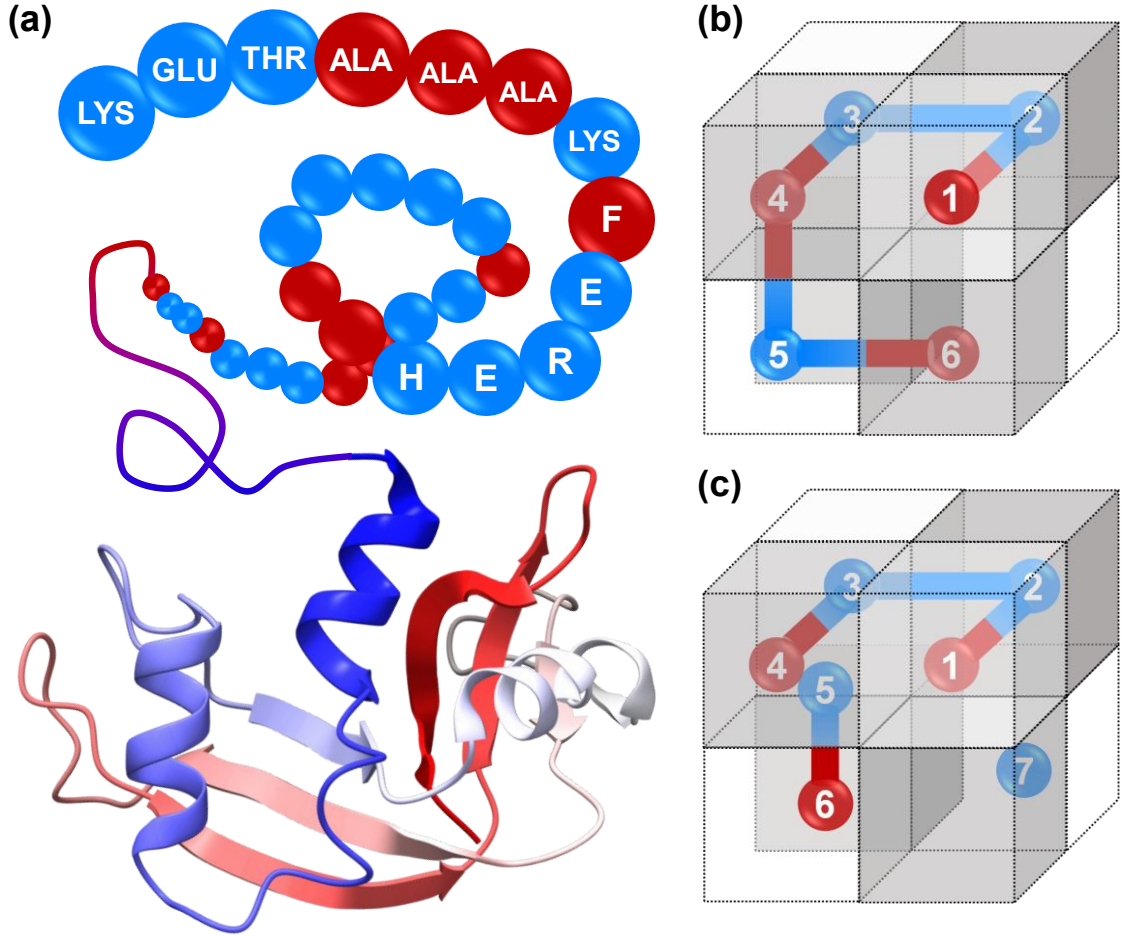

FIG. 1. Schematic representation of the protein folding process and its modeling using the HP model. (a) The folding of bovine pancreatic ribonuclease is determined by its amino acid sequence, where each amino acid is represented as a bead. Hydrophobic residues are depicted as red spheres, while polar residues are shown as blue spheres. (b) In the early folding stages, the polypeptide chain explores various conformations, which may lead to misfolding. (c) The final state represents the stable native conformation of the protein, characterized by most hydrophobic interactions formed for the given amino acid sequence.

## II. METHODS

### A. 3-D Lattice HP Protein Folding Model

This study employed the 3-D HP model to address the protein folding problem [9]. Fig. 1(a) illustrates the folding of bovine pancreatic ribonuclease, whose folding configuration is determined by its amino acid sequence [11]. If folded correctly, these amino acids form a self-avoiding chain in which the amino acid sequence directly affects the folding configuration of the chain. Specifically, when two hydrophobic amino acid molecules are adjacent but not directly connected along the chain, a hydrophobic interaction is considered to form between them. The amino acid chain tends to fold in a manner that maximizes the number of hydrophobic interactions, ultimately determining the overall structure of the protein.

### B. Encoding Protein Folding to Ising Model

To address the protein folding problem using the Ising framework, it is essential to map the problem onto the Ising framework and design the energy function such that the optimal protein folding configuration is encoded into the ground state. Based on the fundamental principles of the previously discussed HP model, configurations with more hydrophobic interactions should correspond to lower energy states in the Ising framework. Therefore, the energy of a given amino acid chain can be defined as:

$$E_{HP} = -N_{HH} \quad (1)$$

where $N_{HH}$ denotes the number of hydrophobic interactions.

The next step is to map the 3-D HP model onto the Ising model. In this study, we map the 3-D HP model by simultaneously encoding both the spatial positions and sequence order of the amino acids. Each node in the Ising framework is designed to capture two layers of information: the spatial position of a specific amino acid and its sequential order within the amino acid chain.

For the 3-D HP model, we assume that the amino acid chain occupies a grid defined by length ($L$), width ($W$), and height ($H$), and therefore, the total number of positions is $L \times W \times H$. To reduce the number of nodes required, the amino acid chain nodes are categorized based on parity (odd or even), and all grid positions are similarly divided into corresponding odd and even categories. Odd and even grid positions are arranged resulting in no two consecutive positions share the same parity. Under this scheme, all odd nodes are assumed to be located at odd grid positions, while even nodes are located at even grid positions. Therefore, each node in the Ising framework can be defined as either an odd node $\sigma_s^f$ or an even node $\sigma_{s'}^{f'}$, where $s$ and $s'$ represents respective spatial paths taken, and $f$ and $f'$ represent the sequence order within the chain. Through this method, the number of nodes in the transformed model can be reduced to $\frac{N \times W \times L \times H}{2}$, optimizing the model's complexity while preserving the essential details required to represent the protein folding problem accurately.

After defining the energy function and successfully mapping the HP model onto the Ising framework, the fundamental energy function of the proposed model can be reformulated as follows:

$$E_{HP} = \sum_{|f-f'|>1} C(h_f, h_{f'}) \sum_{s_{near}} \sigma_s^f \sigma_{s'}^{f'} \quad (2)$$

where $C(h_f, h_{f'}) = 1$ if and only if $h_f = h_{f'} = H$, and $C(h_f, h_{f'}) = 0$ in any other cases. The condition $|f - f'| > 1$ excludes scenarios where two amino acids are directly connected, ensuring that only hydrophobic interactions contribute to lowering the energy. This term is designed to promote hydrophobic interaction by lowering the energy whenever hydrophobic interactions are present. In the second summation, $s_{near}$ denotes all pairs of spatially adjacent nodes.

## C. Penalty Energy Incorporating Many-body Interactions

The fundamental energy function Eq. (2) is insufficient to map the protein folding problem to the Ising framework. This energy function is inherently flawed and can lead to multiple errors, such as multiple amino acids located at a single position or forming fragmented chains rather than a single protein chain in the model, as shown in Fig. 1(b). To address these problems, a penalty term is typically introduced to penalize these erroneous folding configurations, ensuring the correct folding configuration is mapped to the ground state of the Ising framework.

In prior works, conventional approaches to designing the penalty component involved imprecise constraints and support for incorrect folding configurations. These approaches often introduce additional terms into the energy function without fully accounting for the broader energy landscape, resulting in its degradation and reduced model performance. In contrast, this study introduces a penalization approach in which the penalty terms are calculated directly from a truth table-based direct mapping method [27]. This approach ensures that the energy landscape of the penalty component forms a binary step function, thereby significantly reducing the number of energy levels and boosting the overall p-circuit performance.

The design of the penalty function terms is based on turn-based constraints. Specifically, whenever a specific amino acid occupies a particular position, i.e., a given $\sigma = 1$, a corresponding truth table will be generated, which ensures that only one neighbor node is occupied by amino acid. After that, the penalty terms are directly calculated from this truth table [27], as shown in Fig. 2. Besides turn-based constraints, constraints on amino acid positions and sequences are also imposed. These two constraints ensure that each gird is occupied by at most one amino acid, and each amino acid occupies a unique position in the sequence. The expressions for these two constraints are as follows:

$$E_{location} = \sum_{f_1 \neq f_2}\sum_{s} \sigma_s^{f_1} \sigma_s^{f_2} + \sum_{s_1 \neq s_2}\sum_{f} \sigma_{s_1}^{f} \sigma_{s_2}^{f} + \{same\ for\ odd\ parity\}. \quad (3)$$

Owing to the high effectiveness of the directly calculated turn-based constraints, the intensity of other constraints can be kept minimal. This prevents any substantial alteration to the overall energy landscape. Furthermore, as the turn-based energy constraints effectively set the sequence, additional energy constraints are solely required to superimpose further constraints without requiring support terms.

The complete energy landscape is obtained by integrating two components of the penalty function: the turn-based penalty function and the location-based penalty function Eq. (3). The total energy of the model is given by:

$$E_{total} = E_{HP} + \lambda_1 E_{turn} + \lambda_2 E_{location} \quad (4)$$

where $\lambda_1$ and $\lambda_2$ are the parameters that govern the intensity of the penalty terms. The values of these two parameters cannot be set excessively low, as the penalty terms must sufficiently constrain incorrect folding configurations. Meanwhile, their values should be minimized to maintain the simplicity of the overall energy landscape. Despite the energy function Eq. (4) appearing to be complex, the overall energy landscape of the proposed model becomes highly concise. For small-scale problems, the energy levels are roughly half of those produced by the conventional model. The advantage of the proposed design becomes increasingly evident when the problem size increases. Furthermore, this proposed model avoids excessive utilities of many-body interactions, which helps maintain the practicality of the circuit.

### D. Simulation of the Proposed Model Using P-circuit

Once the energy function is determined, the configuration parameters of the proposed model can be directly derived by mapping the terms involving $\sigma$ in the energy expression onto a corresponding p-circuit. With these prepared configuration parameters, an Ising machine can then be constructed with a p-circuit to address protein folding problems, specifically within the framework of the 3-D HP model.

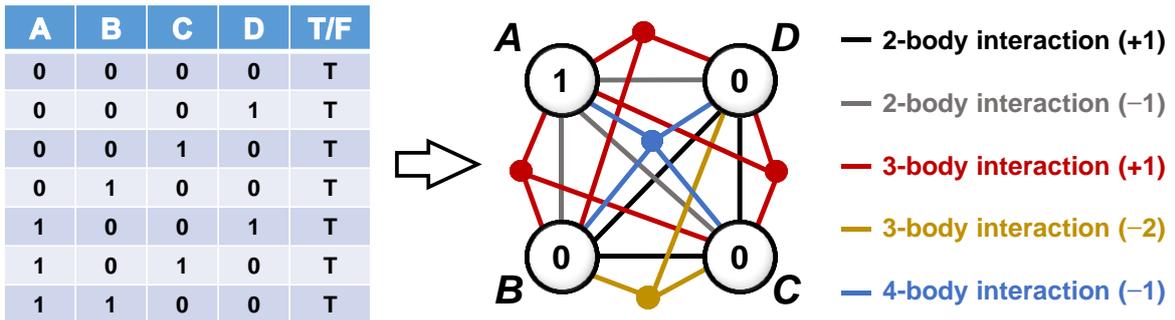

FIG. 2: Schematic of the direct application of a truth table to calculate the optimized turn-based penalty energy in a lattice model. The diagram illustrates the penalty energy of a central node A with three neighboring nodes B, C, and D. The truth table enumerates all configurations where the Boolean function evaluates to true (T), while all other input combinations for nodes A, B, C, and D yield false (F) and are omitted for clarity. The interaction strength between nodes is indicated by colored lines.

In the p-circuit, p-bits function as "spins" nodes for the proposed model, and a resistor network is employed to construct the entire model. There are various types of p-bits available, including MTJ-based devices [28], [29], [30], [31], CMOS-based devices [32], [33], [34], FeFET-based devices [35], [36], and other emerging probabilistic devices [37], [38], [39]. This study does not impose constraints on the type of p-bit used. The only requirement is that the device can provide a probabilistic sigmoidal function according to input current:

$$s_i(t) = \theta\{rand(-1,1) + \tanh[I_i(t)]\} \quad (5)$$

where $\theta(x)$ is a unit step function and it is defined as being zero for all input values less than zero and one for all input values equal to or greater than zero. rand $(-1, +1)$ uniformly distributed between $-1$ and $+1$, and $I_i$ presents the input of the $i^{th}$ p-bit. The connections between nodes are implemented using a resistor network. The complete expression for the input current of each p-bit is given by:

$$I_i(t) = I_0(h_i + \sum_j J_{ij}s_j(t) + \sum_{j,k} K_{ijk}s_j(t)s_k(t) + \sum_{j,k,l} L_{ijkl}s_j(t)s_k(t)s_l(t) + \cdots) \quad (6)$$

where $s$ denotes the binary values, i.e., 0, +1. $I_0$ serves as a global dimensionless parameter governing the overall strength of interactions to each p-bit. Local bias term $h$ is assigned to each node, while $J$, $K$ and $L$ characterize the two-body, three-body and four-body interactions between nodes, respectively.

The construction of the p-circuit is based on the energy function Eq. (4). A passive resistor network is employed for connections involving two-body or fewer (i.e., interconnections with only one or two $s$ variables multiplied together in the energy function). Inspired by using CMOS XNOR gates to implement higher-order interactions for bipolar variables [40], here, for many-body interactions with binary representation of $s$, $M-2$ conventional CMOS AND gates are incorporated at the connection points to achieve the multiplication of p-bit's output voltage, where $M$ represents the order of the many-body interactions. This design enables the p-circuit to implement complex many-body interactions effectively.

At thermal equilibrium, the steady probability for each state configuration can be described by the Boltzmann Law:

$$P(\{s\}) = \frac{exp\left(-\frac{E(s)}{T}\right)}{\sum_{i,j} exp\left(-\frac{E(s)}{T}\right)} \quad (7)$$

where $T$ is the pseudo-temperature parameter that reflects the overall stochasticity of the system. During the iteration, state configurations with lower energy will be emphasized.

## III. RESULTS

In this section, we assess the performance of the proposed model through simulated p-circuit implementations across three representative case studies. First, a 2×2×2 grid is utilized to address a protein folding problem involving an amino acid sequence of length 6. This example highlights the ability of many-body interactions to simplify the energy landscape and facilitate the search for optimal configurations. Subsequently, 4×4 and 4×4×4 grids are employed to solve a protein folding problem with a sequence length of 10 to compare the performance of the 3-D HP model with its 2-D counterpart, which will underscore the advantages of many-body interactions in accurately modeling the folding process. Finally, the proposed model is applied to a more complex protein folding problem with a sequence length of 36, leveraging a 4×4×4 grid to showcase its scalability and efficacy in handling more complicated systems.

### A. Protein folding in a $2 \times 2 \times 2$ grid

First, we consider an amino acid sequence with a length of $N = 6$ (HPPHPH) is folded within a $2 \times 2 \times 2$ grid. The energy functions incorporating up to only two-body interactions are compared with those that include up to four-body interactions. Fig. 3 illustrates that the incorporation of many-body interactions can significantly simplify the energy landscape with the number of accessible energy levels greatly reduced from 229 to 104.

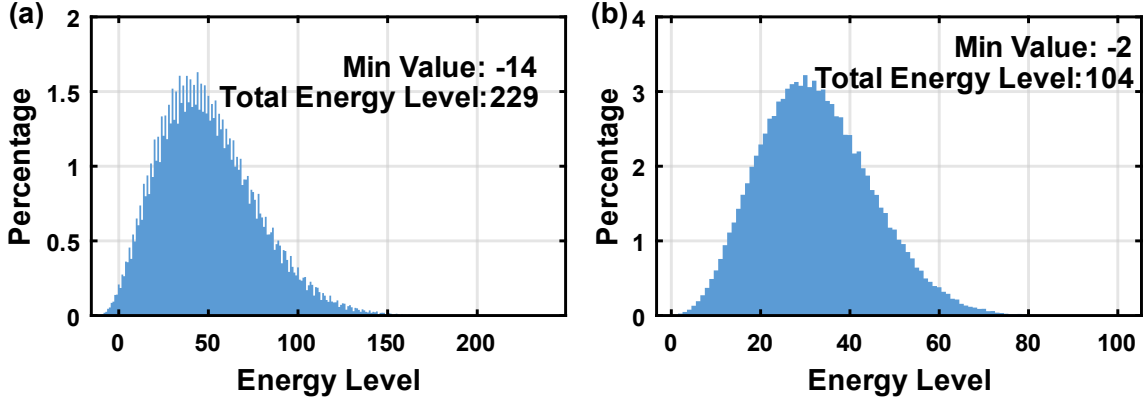

FIG. 3: Theoretical energy level distributions of the proposed model for solving a protein folding problem with a sequence length of $N = 6$ within a $2 \times 2 \times 2$ grid. (a) Energy level distribution of the system utilizes only two-body interactions between neighboring nodes. The minimum energy value achieved is $-14$, with a total energy level of 229. (b) Energy level distribution of the many-body interactions-based system, where the minimum energy value obtained is $-2$, with a total energy level of 104. The x-axis denotes the values of energy levels, and the y-axis indicates the theoretical distribution of energy levels occurring in the solution space.

Furthermore, the number of hydrophobic interactions directly corresponds to the absolute energy values when many-body interactions are utilized, providing a key advantage for performance evaluation and annealing. The additional connection freedom introduced by many-body interactions enables the penalty function to penalize incorrect folding configurations while ensuring that correct configurations remain unaffected. Consequently, no additional support terms are required in the energy function, which could otherwise lower the energy of both correct and incorrect folding configurations. Moreover, the design regulates the order of many-body interactions based on the number of interconnected nodes. As mentioned, turn penalty terms are calculated from the interaction between neighboring turns. In a $2 \times 2 \times 2$ grid, each node can connect to three neighboring nodes at most, resulting in a maximum of four interconnected nodes and hence four-body interactions. In a 3-D grid, each node can connect to up to six neighboring nodes, allowing for a maximum of seven-body interactions. This turn-based design effectively constrains the order of many-body interactions to relatively low values, maintaining the simplicity of the p-circuit.

### B. Comparative Analysis of the Protein Folding Problem Across Different Dimensions and Mapping Methods

Here, we present simulation results from applying our p-circuits to solve the protein folding problem on a small scale. To highlight the specific advantages of the 3-D HP model, the problem is modeled in two ways: a 4×4 2-D HP model and a 4×4×4 3-D HP model. Both models are applied to an amino acid sequence of length $N = 10$. The corresponding results are shown in Fig. 4.

The HP model organizes all amino acids into a grid and assumes that the number of hydrophobic interactions determines protein optimality. As illustrated in Fig. 4(a), the 3-D HP model significantly enhances interaction potential, as each site can engage in hydrophobic interactions with up to six neighboring sites, corresponding to the six possible spatial directions in a 3-D lattice. In contrast, the 2-D HP model confines hydrophobic interactions to four possible directions, restricting the diversity of interaction patterns for amino acid sequences. The additional spatial degrees of freedom in the 3-D HP model enable more diverse and efficient folding solutions compared to the 2-D model. As shown in Fig. 4(b), the 2-D HP model provides solutions with a maximum of three hydrophobic interactions. In comparison, the 3-D HP model identifies configurations with up to four hydrophobic interactions, offering more optimal solutions, as highlighted in Fig. 4(c). Furthermore, the increased variability afforded by the 3-D HP model more accurately represents the intrinsic three-dimensional nature of protein folding observed in real biological systems. This comparison underscores the distinct folding configurations generated by the two models despite utilizing identical folding rules. The enhanced capacity of the 3-D HP model to capture the spatial complexity of protein folding dynamics results in more precise and biologically relevant predictions of protein behavior.

Following the discussion on the enhanced descriptive capability of the 3D HP model, we then analyze the energy evolution during the annealing process of the p-circuit in its search for global minima with the results presented in Fig. 5. Specifically, the performance of two p-circuits is compared: one implementing the proposed many-body mapping method and the other excluding many-body interactions. Both p-circuits are designed to address the protein folding problem within a 4 × 4 × 4 grid containing ten nodes. A linear simulated annealing schedule is used in this study. The increase of $I_0$ corresponds to a global decrease in the resistance value of all interacting resistors in the simulated p-circuit to modulate the overall stochasticity of the p-circuit. At a higher $I_0$, the circuit exhibits better deterministic behavior. It tends to be stable while concurrently increasing the risk of becoming trapped in local energy minimum states. When the system reaches a local optimal state, characterized by an unchanged p-bit state configuration for 50 iterations, the resistance is reset to its initial high value, enabling the system to escape local minima. The simulation result in Fig. 5 highlights the time required for various folding configurations to reach their minimum energy. The p-circuit incorporating many-body interactions with annealing achieves the global minima most efficiently, requiring only 1% of the iterations compared to other scenarios. Meanwhile, the p-circuit employing many-body interactions without annealing performs slightly faster than the configuration with only two-body interactions and annealing. These results underscore that both many-body interactions and annealing play critical roles in improving the efficiency of the p-circuit.

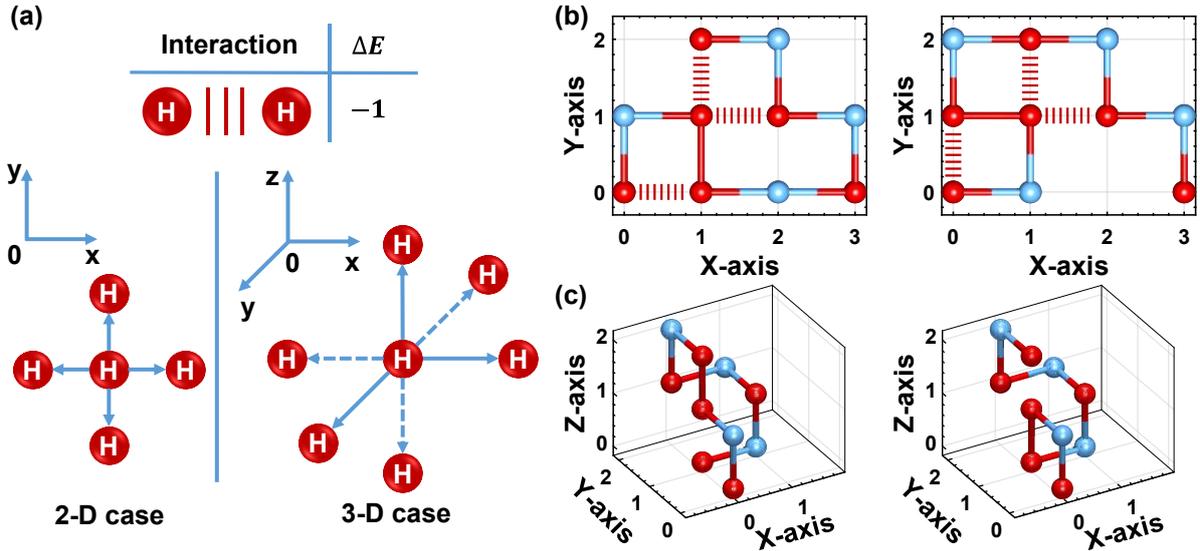

FIG. 4: The 3-D model enhances hydrophobic interaction potential compared to its 2-D counterpart, as it applies to a 3-D orientation where the interaction energy ΔE between non-adjacent hydrophobic residues (H) is −1 owing to the formation of hydrophobic interactions. In contrast, adjacent hydrophobic residues do not form interactions. (b) Optimal folding path solved by the 2-D HP model for a sequence of length $N = 10$. (c) Optimal folding path obtained by the 3-D HP model for the same $N = 10$ sequence.

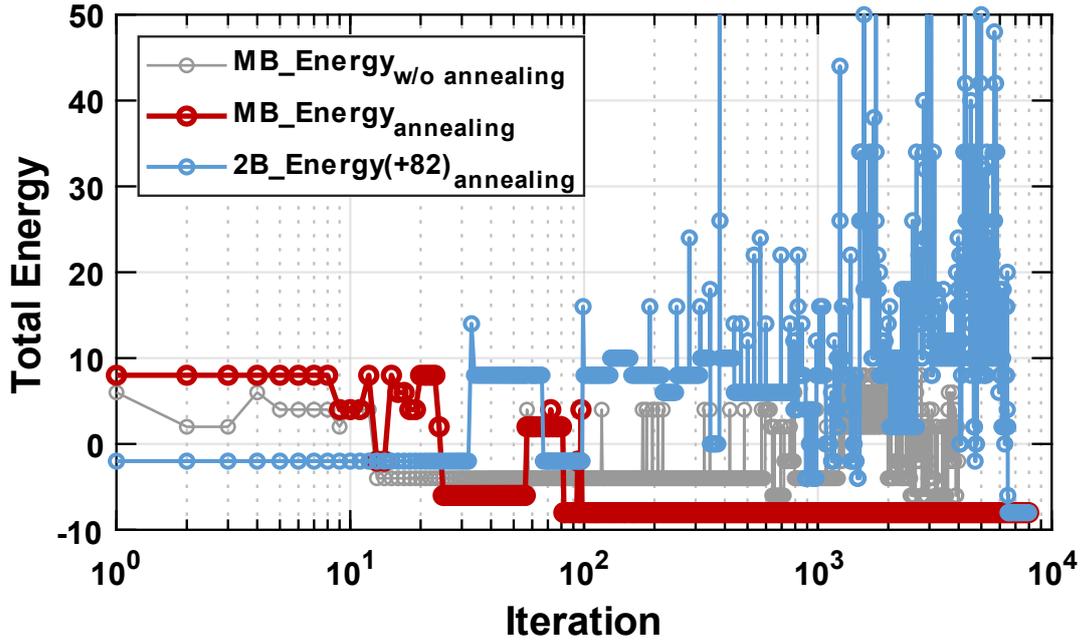

FIG. 5: Real-time fluctuation of the system energy during the evolution. Red curve: a simulated p-circuit utilizing many-body interactions and updated with simulated annealing. Blue curve: a simulated p-circuit utilizing only two-body interactions and updated with simulated annealing, in which the energy value is offset by a constant value of 82 (denoted as E') to align its energy minima to the other two scenarios. Gray curve: a simulated p-circuit with many-body interactions but without simulated annealing. The X-axis is presented on a logarithmic scale.

The performance improvement of the p-circuit utilizing many-body interactions is attributed to the enhanced flexibility introduced by these interactions. These increased degrees of freedom facilitate a more precise penalty function formulation, resulting in a simplified and more regular energy landscape, thus enabling the p-circuit to search for the lowest energy state more effectively. Simultaneously, the enhancement achieved through simulated annealing arises from its ability to modulate the stochastic behavior of the p-circuit. This modulation allows the circuit to search for minimum energy states while retaining sufficient flexibility to escape local minima. Furthermore, the results indicate that the integration of many-body interactions with annealing could offer optimal performance, as these two mechanisms function through distinct improvement pathways. This is supported by the observation that the p-circuit employing both many-body interactions and annealing consistently achieves the global minimum energy state in the shortest time.

## C. Large-Scale Simulation Results of the 3-D HP Model

Then, we examine a more complex protein folding problem involving 36 nodes, modeled within a 4×4×4 grid. For this case, the p-circuit leverages many-body interactions and is optimized using a simulated annealing approach. The simulated annealing method employed here introduces modifications compared to the previous example, incorporating a linear reduction in resistance combined with abrupt resistance adjustments. Given the scale of the problem, with each simulation requiring millions of iterations, the p-circuit is designed to evaluate the solution dynamically. A solution is deemed viable and stored only if the entire p-circuit configuration remains unchanged for 50 consecutive iterations; otherwise, it is discarded. The results of these simulations are presented in Fig. 6. Addressing this large-scale problem required the construction of a network comprising 1154 p-bits. Despite the significant size of the p-circuit, the carefully designed energy function ensures that the annealing process operates efficiently and is completed within a practical timeframe. Over the course of 800 thousand annealing cycles, the p-circuit identified 2534 potential solutions.

This performance surpasses other approaches employing non-conventional computing, which cannot handle problems of similar complexity [24]. The identified optimal solution includes 18 hydrophobic interactions, further validated through benchmarking against previous work [17]. Moreover, the model uncovers not only the optimal protein folding configuration but also several viable alternatives. Notably, most potential solutions are physically reasonable, forming complete chains without fragmentation or instances of multiple amino acids occupying a single node. This proves the efficacy of the penalty term based on many-body interactions in penalizing incorrect configurations, meanwhile underscoring its effectiveness in exploring multiple solutions, thereby contributing to a deeper understanding of potential protein folding paths. Furthermore, the high proportion of reasonable solutions among those stored significantly reduces the computational burden of the post-processing algorithm, enabling faster determination of optimal solutions. These findings highlight the effectiveness of the p-circuit, augmented by many-body interactions, in facilitating the search for optimal configurations in large-scale protein folding problems.

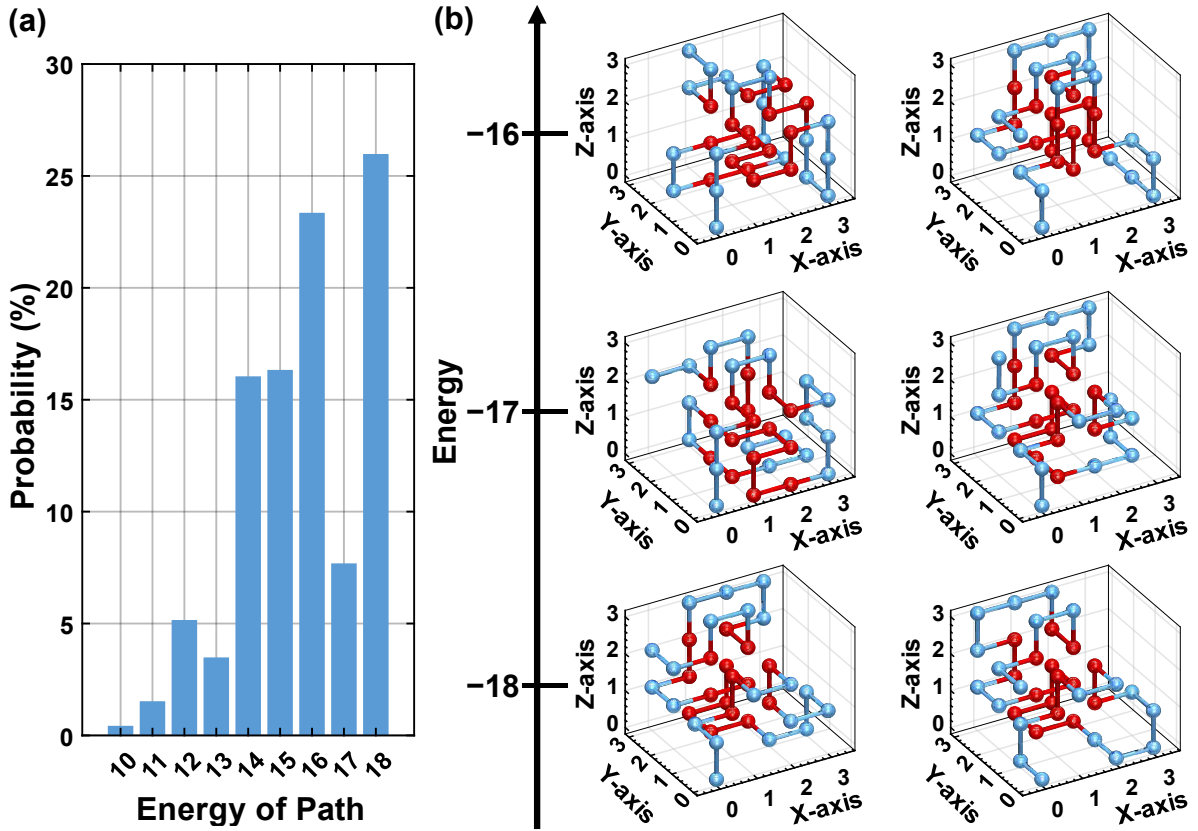

FIG. 6: Simulation results for a protein folding problem with a sequence length of $N = 36$ within a $4 \times 4 \times 4$ grid. A total of 800 thousand iterations were performed in the simulation. (a) Distribution of the energy levels for all solved folding paths, with a total of 2534 distinct paths identified. (b) Representative examples of possible protein folding configurations obtained by the p-circuit. Red spheres represent hydrophobic amino acids, while blue spheres represent polar amino acids. The x-axis represents the absolute value of the energy level of the folding path.

## IV. CONCLUSION

This study introduces a novel approach to addressing the protein folding problem by integrating two key innovations: mapping the 3-D HP model onto an Ising model and solving it using a p-circuit based on p-bits. These innovations synergistically address critical challenges in protein folding computations. The mapping technique refines the translation of the 3-D HP model onto an Ising framework by simplifying the energy landscape, resulting in a more compact and regular energy landscape. This refinement enhances the efficiency of Ising machines, enabling rapid identification of optimal configurations through linear simulated annealing. Concurrently, the use of p-bit-based probabilistic circuits provides a scalable and practical alternative to quantum computing, overcoming limitations such as the need for cryogenic conditions and limited problem sizes. Simulations demonstrate the effectiveness of our p-circuits in solving protein folding problems with sequence lengths up to $N$=36 within a 4×4×4 grid. Applying the method outlined in this study, 544 distinct folding configurations with 18 hydrophilic interactions were identified. This performance surpasses existing non-conventional computing approaches in terms of problem scale and solution quality.

Beyond its immediate application, the proposed model offers significant versatility and adaptability. By mapping the 3-D HP model onto the Ising framework, the method positions itself for integration with quantum computing platforms, extending its applicability beyond probabilistic computing. Furthermore, the model can be generalized to incorporate more sophisticated representations, such as the Miyazawa-Jernigan model, by integrating detailed energy intensity coefficients to capture complex hydrophobic and hydrophilic interactions. This flexibility broadens the model's utility, enabling it to address a broader range of protein folding problems with enhanced accuracy and efficiency. The approach thus represents a reliable and scalable tool for advanced research in computational protein science and contributes to the development of next-generation non-conventional computing frameworks.

ACKNOWLEDGMENTS

This work at the National University of Singapore was supported by FRC-A-8000194-01-00. G.C. L. would also like to thank the financial support from the National Science and Technology Council (NSTC) under grant number NSTC 112-2112-M-A49-047-MY3, the Co-creation Platform of the Industry-Academia Innovation School, NYCU, under the framework of the National Key Fields Industry-University Cooperation and Skilled Personnel Training Act, and the Advanced Semiconductor Technology Research Center from The Featured Areas Research Center Program within the framework of the Higher Education Sprout Project by the Ministry of Education (MOE) in Taiwan.

**References**

[1] K. A. Dill and J. L. MacCallum, "The Protein-Folding Problem, 50 Years On," *Science*, vol. 338, no. 6110, pp. 1042–1046, Nov. 2012, doi: 10.1126/science.1219021.

[2] M. Wang and R. J. Kaufman, "The impact of the endoplasmic reticulum protein-folding environment on cancer development," *Nat. Rev. Cancer*, vol. 14, no. 9, pp. 581–597, Sep. 2014, doi: 10.1038/nrc3800.

[3] M. G. Krone et al., "Effects of Familial Alzheimer's Disease Mutations on the Folding Nucleation of the Amyloid β-Protein," *J. Mol. Biol.*, vol. 381, no. 1, pp. 221–228, Aug. 2008, doi: 10.1016/j.jmb.2008.05.069.

[4] W. Poewe et al., "Parkinson disease," *Nat. Rev. Dis. Primer*, vol. 3, no. 1, p. 17013, Mar. 2017, doi: 10.1038/nrdp.2017.13.

[5] M. V. Khan, S. M. Zakariya, and R. H. Khan, "Protein folding, misfolding and aggregation: A tale of constructive to destructive assembly," *Int. J. Biol. Macromol.*, vol. 112, pp. 217–229, Jun. 2018, doi: 10.1016/j.ijbiomac.2018.01.099.

[6] D. J. Selkoe, "Folding proteins in fatal ways," *Nature*, vol. 426, no. 6968, pp. 900–904, Dec. 2003, doi: 10.1038/nature02264.

[7] J. Jumper et al., "Highly accurate protein structure prediction with AlphaFold," *Nature*, vol. 596, no. 7873, pp. 583–589, Aug. 2021, doi: 10.1038/s41586-021-03819-2.

[8] A. P. AhYoung, A. Koehl, D. Cascio, and P. F. Egea, "Structural mapping of the C lp B ATP ases of *Plasmodium falciparum* : Targeting protein folding and secretion for antimalarial drug design," *Protein Sci.*, vol. 24, no. 9, pp. 1508–1520, Sep. 2015, doi: 10.1002/pro.2739.

[9] K. F. Lau and K. A. Dill, "A lattice statistical mechanics model of the conformational and sequence spaces of proteins," *Macromolecules*, vol. 22, no. 10, pp. 3986–3997, Oct. 1989, doi: 10.1021/ma00200a030.

[10] K. A. Dill et al., "Principles of protein folding — A perspective from simple exact models," *Protein Sci.*, vol. 4, no. 4, pp. 561–602, Apr. 1995, doi: 10.1002/pro.5560040401.

[11] C. B. Anfinsen, "Principles that Govern the Folding of Protein Chains," *Science*, vol. 181, no. 4096, pp. 223–230, Jul. 1973, doi: 10.1126/science.181.4096.223.

[12] C. B. Anfinsen, E. Haber, M. Sela, and F. H. White, "THE KINETICS OF FORMATION OF NATIVE RIBONUCLEASE DURING OXIDATION OF THE REDUCED POLYPEPTIDE CHAIN," *Proc. Natl. Acad. Sci.*, vol. 47, no. 9, pp. 1309–1314, Sep. 1961, doi: 10.1073/pnas.47.9.1309.

[13] B. Berger and T. Leighton, "Protein Folding in the Hydrophobic-Hydrophilic ( *HP* ) Model is NP-Complete," *J. Comput. Biol.*, vol. 5, no. 1, pp. 27–40, Jan. 1998, doi: 10.1089/cmb.1998.5.27.

[14] P. Crescenzi, D. Goldman, C. Papadimitriou, A. Piccolboni, and M. Yannakakis, "On the Complexity of Protein Folding," *J. Comput. Biol.*, vol. 5, no. 3, pp. 423–465, Jan. 1998, doi: 10.1089/cmb.1998.5.423.

[15] C.-J. Lin and M.-H. Hsieh, "An efficient hybrid Taguchi-genetic algorithm for protein folding simulation," *Expert Syst. Appl.*, vol. 36, no. 10, pp. 12446–12453, Dec. 2009, doi: 10.1016/j.eswa.2009.04.074.

[16] A. Shmygelska, R. Aguirre-Hernández, and H. H. Hoos, "An Ant Colony Optimization Algorithm for the 2D HP Protein Folding Problem," in *Ant Algorithms*, vol. 2463, M. Dorigo, G. Di Caro, and M. Sampels, Eds., in Lecture Notes in Computer Science, vol. 2463. , Berlin, Heidelberg: Springer Berlin Heidelberg, 2002, pp. 40–52. doi: 10.1007/3-540-45724-0_4.

[17] N. Boumedine and S. Bouroubi, "Protein folding in 3D lattice HP model using a combining cuckoo search with the Hill-Climbing algorithms," *Appl. Soft Comput.*, vol. 119, p. 108564, Apr. 2022, doi: 10.1016/j.asoc.2022.108564.

[18] S. Pal, M. Bhattacharya, S.-S. Lee, and C. Chakraborty, "Quantum Computing in the Next-Generation Computational Biology Landscape: From Protein Folding to Molecular Dynamics," *Mol. Biotechnol.*, vol. 66, no. 2, pp. 163–178, Feb. 2024, doi: 10.1007/s12033-023-00765-4.

[19] P. Chandarana, N. N. Hegade, I. Montalban, E. Solano, and X. Chen, "Digitized Counterdiabatic Quantum Algorithm for Protein Folding," *Phys. Rev. Appl.*, vol. 20, no. 1, p. 014024, Jul. 2023, doi: 10.1103/PhysRevApplied.20.014024.

[20] A. Perdomo, C. Truncik, I. Tubert-Brohman, G. Rose, and A. Aspuru-Guzik, "On the construction of model Hamiltonians for adiabatic quantum computation and its application to finding low energy conformations of lattice protein models," *Phys. Rev. A*, vol. 78, no. 1, p. 012320, Jul. 2008, doi: 10.1103/PhysRevA.78.012320.

[21] A. Perdomo-Ortiz, N. Dickson, M. Drew-Brook, G. Rose, and A. Aspuru-Guzik, "Finding low-energy conformations of lattice protein models by quantum annealing," *Sci. Rep.*, vol. 2, no. 1, p. 571, Aug. 2012, doi: 10.1038/srep00571.

[22] M. Fingerhuth, T. Babej, and C. Ing, "A quantum alternating operator ansatz with hard and soft constraints for lattice protein folding," Oct. 31, 2018, *arXiv*: arXiv:1810.13411. Accessed: Aug. 15, 2024. [Online]. Available: http://arxiv.org/abs/1810.13411

[23] A. Robert, P. Kl. Barkoutsos, S. Woerner, and I. Tavernelli, "Resource-efficient quantum algorithm for protein folding," *Npj Quantum Inf.*, vol. 7, no. 1, p. 38, Feb. 2021, doi: 10.1038/s41534-021-00368-4.

[24] A. Irbäck, L. Knuthson, S. Mohanty, and C. Peterson, "Folding lattice proteins with quantum annealing," *Phys. Rev. Res.*, vol. 4, no. 4, p. 043013, Oct. 2022, doi: 10.1103/PhysRevResearch.4.043013.

[25] A. Irbäck, L. Knuthson, S. Mohanty, and C. Peterson, "Using quantum annealing to design lattice proteins," *Phys. Rev. Res.*, vol. 6, no. 1, p. 013162, Feb. 2024, doi: 10.1103/PhysRevResearch.6.013162.

[26] K. Y. Camsari et al., "From Charge to Spin and Spin to Charge: Stochastic Magnets for Probabilistic Switching," *Proc. IEEE*, vol. 108, no. 8, pp. 1322–1337, Aug. 2020, doi: 10.1109/JPROC.2020.2966925.

[27] Y. He, S. Luo, C. Fang, and G. Liang, "Direct design of ground-state probabilistic logic using many-body interactions for probabilistic computing," *Sci. Rep.*, vol. 14, no. 1, p. 15076, Jul. 2024, doi: 10.1038/s41598-024-65676-z.

[28] K. Y. Camsari, R. Faria, B. M. Sutton, and S. Datta, "Stochastic p-Bits for Invertible Logic," *Phys. Rev. X*, vol. 7, no. 3, p. 031014, Jul. 2017, doi: 10.1103/PhysRevX.7.031014.

[29] W. A. Borders, A. Z. Pervaiz, S. Fukami, K. Y. Camsari, H. Ohno, and S. Datta, "Integer factorization using stochastic magnetic tunnel junctions," *Nature*, vol. 573, no. 7774, pp. 390–393, Sep. 2019, doi: 10.1038/s41586-019-1557-9.

[30] K. Y. Camsari, S. Salahuddin, and S. Datta, "Implementing p-bits With Embedded MTJ," *IEEE Electron Device Lett.*, vol. 38, no. 12, pp. 1767–1770, Dec. 2017, doi: 10.1109/LED.2017.2768321.

[31] B. Cai et al., "Unconventional computing based on magnetic tunnel junction," *Appl. Phys. A*, vol. 129, no. 4, p. 236, Apr. 2023, doi: 10.1007/s00339-022-06365-4.

[32] S. C. Smithson, N. Onizawa, B. H. Meyer, W. J. Gross, and T. Hanyu, "Efficient CMOS Invertible Logic Using Stochastic Computing," *IEEE Trans. Circuits Syst. Regul. Pap.*, vol. 66, no. 6, pp. 2263–2274, Jun. 2019, doi: 10.1109/TCSI.2018.2889732.

[33] S. Nikhar, S. Kannan, N. A. Aadit, S. Chowdhury, and K. Y. Camsari, "All-to-all reconfigurability with sparse and higher-order Ising machines," *Nat. Commun.*, vol. 15, no. 1, p. 8977, Oct. 2024, doi: 10.1038/s41467-024-53270-w.

[34] N. S. Singh et al., "CMOS plus stochastic nanomagnets enabling heterogeneous computers for probabilistic inference and learning," *Nat. Commun.*, vol. 15, no. 1, p. 2685, Mar. 2024, doi: 10.1038/s41467-024-46645-6.

[35] S. Luo, Y. He, B. Cai, X. Gong, and G. Liang, "Probabilistic-Bits Based on Ferroelectric Field-Effect Transistors for Probabilistic Computing," *IEEE Electron Device Lett.*, vol. 44, no. 8, pp. 1356–1359, Aug. 2023, doi: 10.1109/LED.2023.3285525.

[36] S. Luo, Y. He, C. Fang, B. Cai, X. Gong, and G. Liang, "The stochastic ferroelectric field-effect transistors-based probabilistic-bits: from device physics analysis to invertible logic applications," *Jpn. J. Appl. Phys.*, vol. 63, no. 2, p. 02SP77, Feb. 2024, doi: 10.35848/1347-4065/ad1bbc.

[37] P. Debashis, V. Ostwal, R. Faria, S. Datta, J. Appenzeller, and Z. Chen, "Hardware implementation of Bayesian network building blocks with stochastic spintronic devices," *Sci. Rep.*, vol. 10, no. 1, p. 16002, Sep. 2020, doi: 10.1038/s41598-020-72842-6.

[38] N. A. Aadit, A. Grimaldi, G. Finocchio, and K. Y. Camsari, "Physics-inspired Ising Computing with Ring Oscillator Activated p-bits," in *2022 IEEE 22nd International Conference on Nanotechnology (NANO)*, Palma de Mallorca, Spain: IEEE, Jul. 2022, pp. 393–396. doi: 10.1109/NANO54668.2022.9928681.

[39] J. Kim et al., "Fully CMOS-Based p-Bits with a Bistable Resistor for Probabilistic Computing," *Adv. Funct. Mater.*, vol. 34, no. 22, p. 2307935, May 2024, doi: 10.1002/adfm.202307935.

[40] Y. He, C. Fang, S. Luo, and G. Liang, "Many-Body Effects-Based Invertible Logic With a Simple Energy Landscape and High Accuracy," *IEEE J. Explor. Solid-State Comput. Devices Circuits*, vol. 9, no. 2, pp. 83–91, Dec. 2023, doi: 10.1109/JXCDC.2023.3320230.